\documentclass[sigconf]{acmart}
\usepackage{algorithm}
\usepackage{algpseudocode}
\usepackage{tabu}
\usepackage{array}
\usepackage{tikz}
\usetikzlibrary{matrix, positioning}
\AtBeginDocument{%
  }

\setcopyright{acmlicensed}
\copyrightyear{2018}
\acmYear{2018}
\acmDOI{XXXXXXX.XXXXXXX}
\acmConference[Conference acronym 'XX]{Companion Proceedings of the ACM Web Conference 2026}{June 03--05,
  2018}{Woodstock, NY}
\acmISBN{978-1-4503-XXXX-X/2018/06}

\begin{document}

\title{FITRep: Attention-Guided Item Representation via MLLMs}

\author{Guoxiao Zhang}
\authornote{Both authors contributed equally to this research.}
\email{zhangguoxiao@meituan.com}
\email{liao27@meituan.com}
\orcid{0009-0004-2628-7094}
\author{Ao Li}
\authornotemark[1]
\affiliation{%
  \institution{Meituan}
  \city{Beijing}
  \country{China}}

\author{Tan Qu}
\authornotemark[1]

\affiliation{%
  \institution{Meituan}
  \city{Beijing}
  \country{China}
}\email{qutan@meituan.com}

\author{Qianlong Xie}
\affiliation{%
  \institution{Meituan}
  \city{Beijing}
  \country{China}}
  \email{xieqianlong@meituan.com}

\author{Xingxing Wang}
\affiliation{%
  \institution{Meituan}
  \city{Beijing}
  \country{China}}
  \email{wangxingxing04@meituan.com}

\renewcommand{\shortauthors}{Guoxiao Zhang, Tan Qu, Ao Li, Qiang Liu, and Xingxing Wang}

\begin{abstract}
Online platforms usually suffer from user experience degradation due to near-duplicate items with similar visuals and text. While Multimodal Large Language Models (MLLMs) enable multimodal embedding, existing methods treat representations as black boxes, ignoring structural relationships (e.g., primary vs. auxiliary elements), leading to \textbf{local structural collapse problem}. To address this, inspired by Feature Integration Theory (FIT), we propose \textbf{FITRep}, the first \textit{attention-guided, white-box item representation} framework for fine-grained \textit{item deduplication}. \textbf{FITRep} consists of: (1) \textbf{Concept-Hierarchical Information Extraction} (CHIE), using MLLMs to extract hierarchical semantic concepts; (2) \textbf{Structure-Preserving Dimensionality Reduction} (SPDR), an adaptive UMAP-based method for efficient information compression; and (3) \textbf{FAISS-Based Clustering} (FBC), a FAISS-based clustering that  assigns each item a
unique cluster id using FAISS. Deployed on Meituan's advertising system, \textbf{FITRep} achieves +3.60\% CTR and +4.25\% CPM gains in online A/B tests, demonstrating both effectiveness and real-world impact.

\end{abstract}

\ccsdesc[500]{Information systems~Information retrieval}

\keywords{Multimodal, Large Language Models, Dimensionality Reduction}

\maketitle

\section{Introduction}
\begin{figure}
    \centering
    \includegraphics[width=\linewidth]{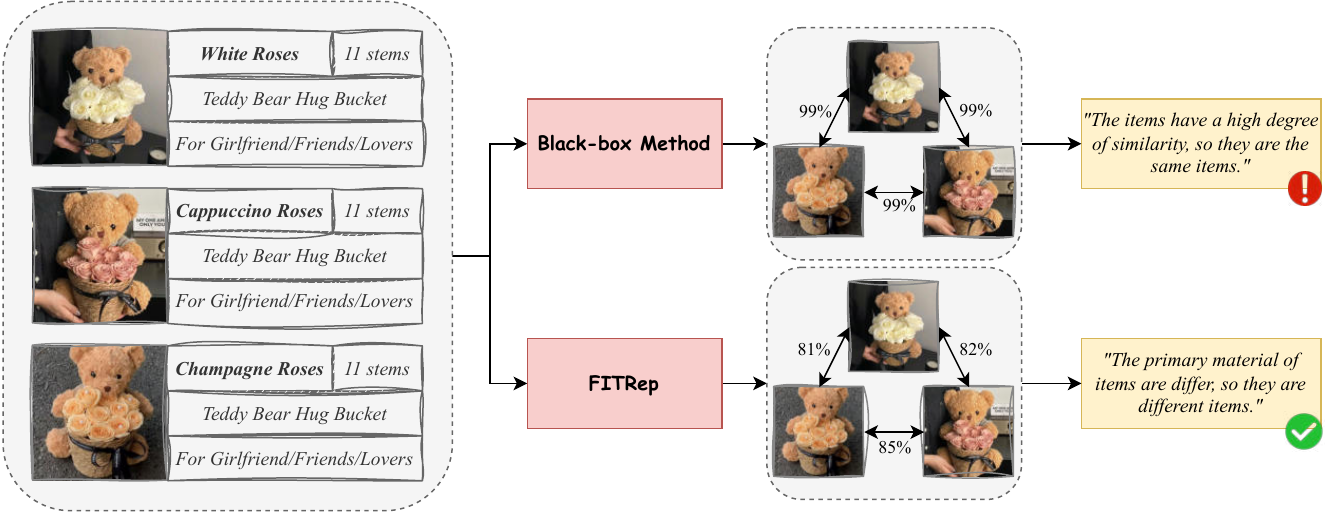}
    \caption{
    An illustration of the \textbf{Local structural collapse problem}.
  }
  \label{fig:case}
\end{figure}

Most online platforms leverage multimodal information such as images and textual titles to enhance user engagement~\cite{wu2022mm}. However, the repeated display of items with near-identical visual and textual content degrades user experience by introducing redundancy. To mitigate this, we propose a multimodal similarity metric for effective \emph{item deduplication} task that identifies and removes redundant or near-duplicate items. Effective \emph{item deduplication} critically depends on the quality of multimodal item representations.

Recently, Multimodal Large Language Models (MLLMs) have shown impressive capabilities in multimodal understanding and generating~\cite{bai2025qwen2, achiam2023gpt}. Existing approaches for obtaining item embeddings via MLLMs broadly fall into two categories: (1) \textit{prompt-based methods}~\cite{ye2025harnessing,zhou2025large}, where simple prompts are employed  to generate image summaries which are transformed into latent vectors using encoding models like BERT~\cite{devlin2019bert}; and (2) \textit{end-to-end joint learning}(NoteLLM-2 ~\cite{zhang2025notellm}), where lightweight MLLMs are fine-tuned exploiting co-occurrence items for better multimodal representation. We refer to such paradigms as \textit{black-box coarse-grained item representation}, which suffers from \textbf{local structural collapse problem}: item images and texts inherently contain structural relationships, such as objective/subjective, form/content, primary/auxiliary materials, etc. Those black-box approaches, lacking explicit modeling of these structures, produce representations that may lose these structural relationships. As an example shown in Figure~\ref{fig:case}, due to the absence of attention mechanisms that differentiate primary from auxiliary materials, items with different primary materials but highly similar auxiliary materials receive spuriously high multimodal similarity scores, leading to false-positive duplicates.

 Inspired by Feature Integration Theory (FIT)~\cite{treisman1980feature}, which suggests that the brain processes basic visual features automatically and separately in a pre-attentive stage, and then combines them accurately through focused attention during a subsequent attentive stage, we propose a holistic framework for \textit{attention-guided, white-box item representation} that includes: (1) \textbf{Concept-Hierarchical Information Extraction} (CHIE) using MLLMs; (2) \textbf{Structure-Preserving Dimensionality Reduction} (SPDR) via adaptive UMAP, and (3) \textbf{FAISS-Based Clustering} (FBC) that assigns differentiated weights to distinct elements based on their attention coefficients, thereby generating the final item representation for FAISS based clustering.

This paper makes three key contributions:  
(1) We propose \textbf{FITRep}, the first \textit{attention-guided, white-box item representation} framework that leverages MLLMs for interpretable and fine-grained perception.  
(2) \textbf{FITRep} comprises three core components: \textbf{CHIE} using MLLMs for high-efficiency information extraction, \textbf{SPDR} via adaptive UMAP for efficient information compression, and \textbf{FBC} for scalable duplicate detection. (3) We deploy \textbf{FITRep} on Meituan's advertising system for \textit{item deduplication} and \textit{CTR prediction}; online A/B tests demonstrate significant improvements of \textbf{+3.60\% in CTR} and \textbf{+4.25\% in CPM}, validating its real-world effectiveness.

\section{Methodology}
\begin{figure*}[t]
    \centering
    \scalebox{0.85}{
    \includegraphics[width=\linewidth]{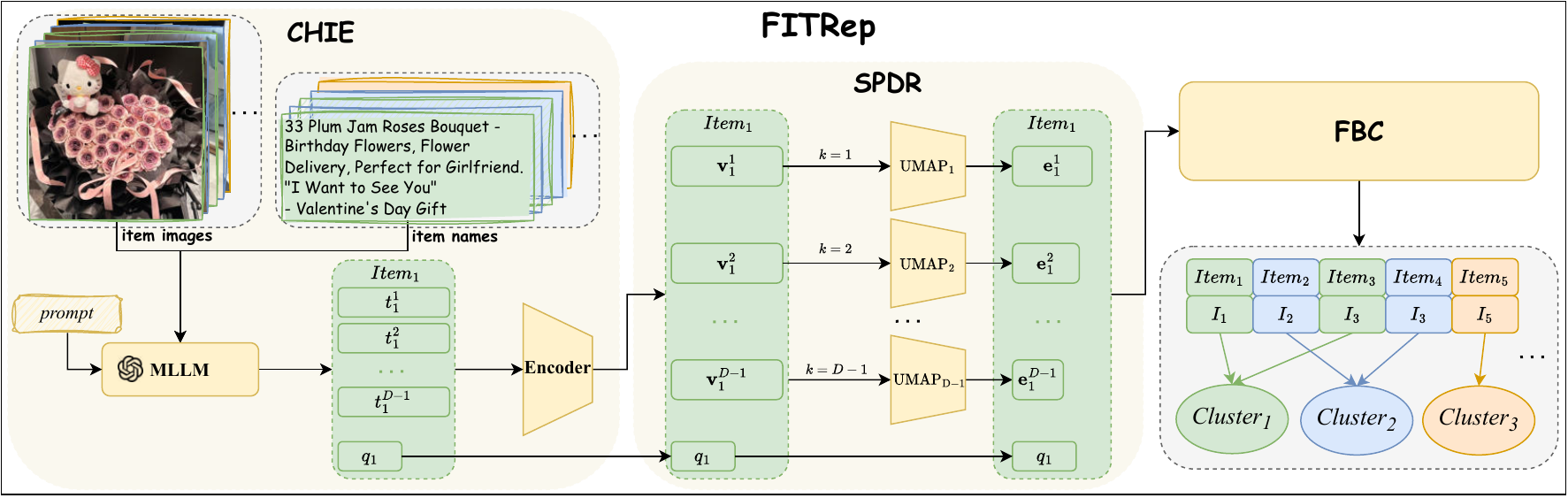}
    }
    \caption{{Overview of our proposed method.}}
    \label{fig:pipeline}
    \vspace{0mm}
    \Description{Overview diagram of the proposed method.}
\end{figure*}

Our framework, illustrated in Figure~\ref{fig:pipeline} includes: (1) \textbf{CHIE} using MLLMs; (2) \textbf{SPDR} via adaptive UMAP; (3) \textbf{FBC} for scalable duplicate detection. Further details are provided in the following section.

\begin{figure}[h]
    \centering
    \includegraphics[width=\linewidth]{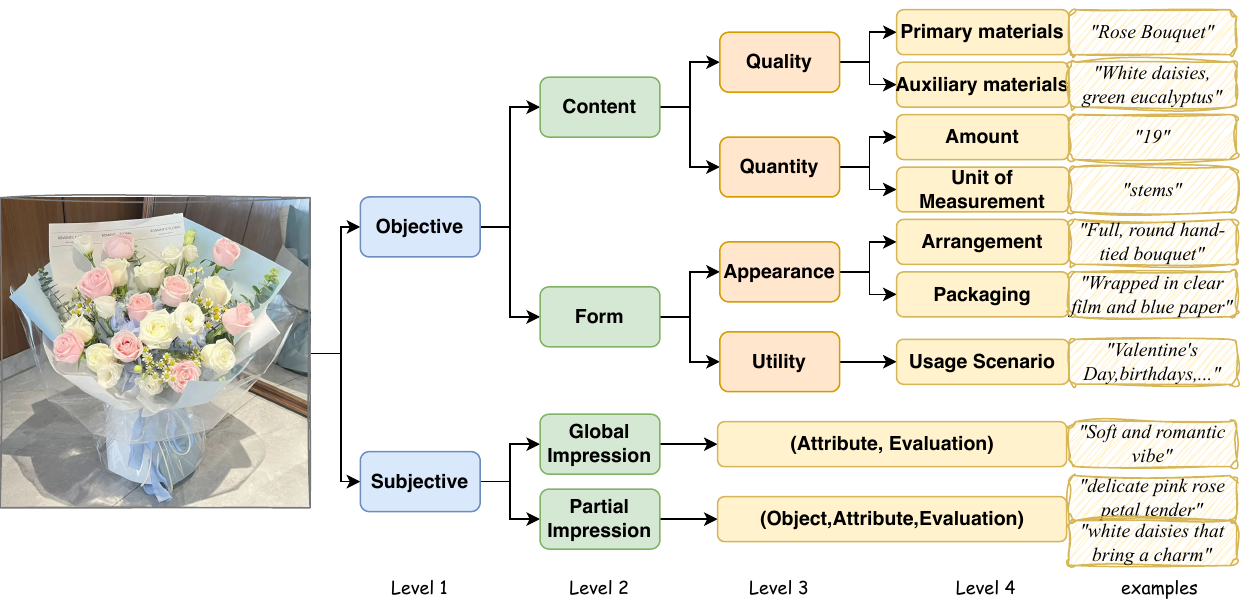}
    \caption{
    An illustration of the \emph{concept hierarchies}.
  }
  \label{fig:concept}
\end{figure}

\subsection{Concept-Hierarchical Information Extraction (CHIE)}
Given that item images and text inherently exhibit structural relationships, we begin by categorizing the multidimensional item information into four hierarchical concept levels (Levels 1–4), as depicted in Figure \ref{fig:concept}. Among these, Level 4 represents the finest granularity of item representation, consisting of $D$ distinct dimensions. In this study, we set $D=8$, as we typically merge the \emph{Amount} and \emph{Unit of Measurement} dimensions into a single dimension \emph{Quantity} for practical purposes.

\begin{figure}[h]
    \centering
    \includegraphics[width=\linewidth]{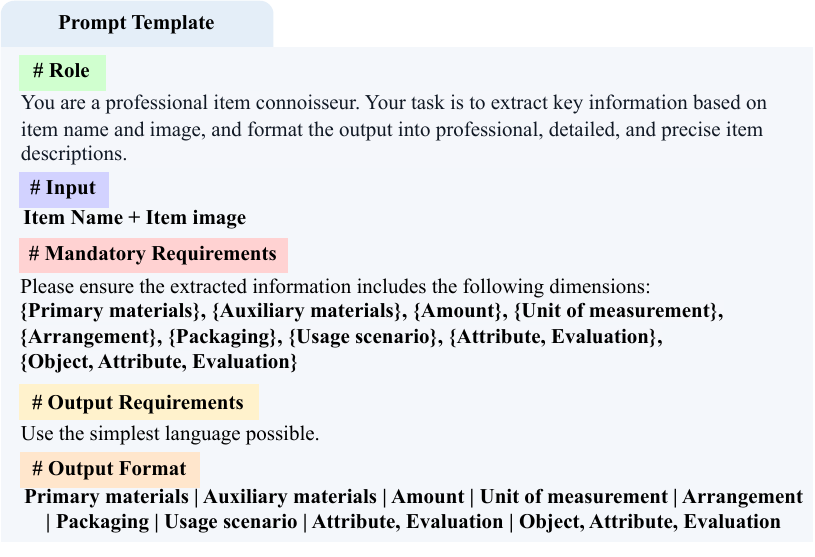}
    \caption{Example of a structured prompt for \textbf{FITRep}.}
  \label{fig:prompt}
\end{figure}

Concurrently, we design a structured prompt that integrates textual and visual inputs to guide the MLLM in extracting dimension-specific item representations (see Figure~\ref{fig:prompt}). Then we utilize a pre-trained Text Encoder \cite{li2023general} to extract dense vector representations from the dimension-specific textual descriptions:

\begin{equation}
    \mathbf{v}^{k} = \mathrm{Encoder}\left({t}^{k}\right), \quad  k=1,\dots,D-1.
\label{eq:1}
\end{equation}

Where $t^k$ represents MLLM generated textual descriptions in the $k$-th conceptual dimension (excluding quantity), we directly use the numerical quantity dimension without representation extraction.

\begin{figure}[ht]
    \centering
    \includegraphics[width=1.0\linewidth]{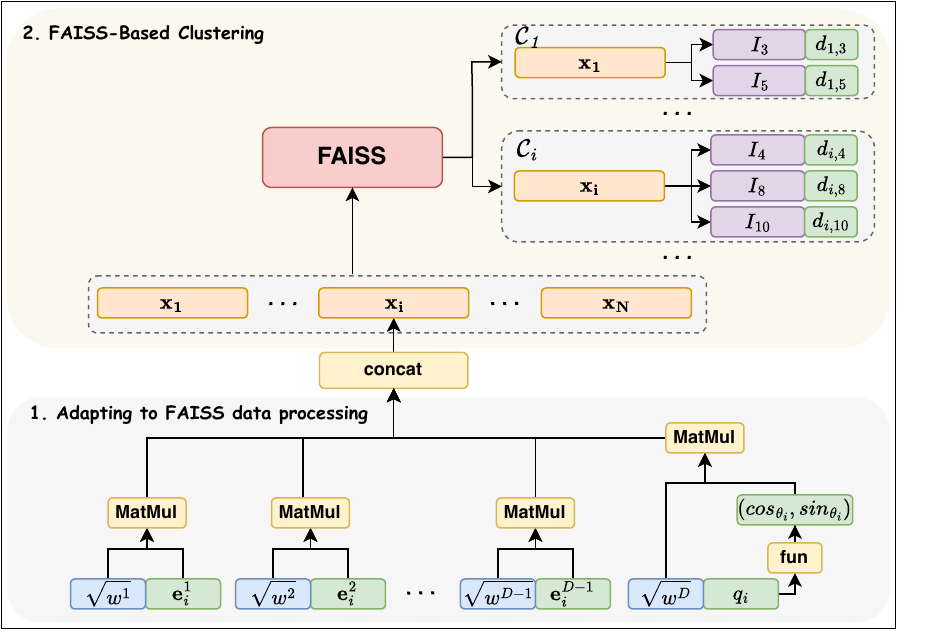}
    \caption{Architecture of the FBC.}
    \label{fig:FBC}
    \vspace{0mm}
\end{figure}

\subsection{Structure-Preserving Dimensionality Reduction (SPDR)}
To reduce high-dimensional embeddings extracted from Equation \eqref{eq:1} while preserving semantic structure, we apply an adaptive parameterized UMAP \cite{sainburg2021parametric}, which dynamically adjusts output dimensionality for different conceptual dimension:

\begin{equation}
    \mathbf{e}^{k} = \mathrm{{UMAP}}\left(k, \mathbf{v}^{k}\right), \quad  k=1,\dots,D-1.
\end{equation}

To facilitate subsequent computations, we represent the embeddings $\mathbf{e}^k$ in their $L_2$-normalized form $\|\mathbf{e}^k\|_2$.

\subsection{FAISS Based Clustering (FBC)}
To enable efficient online deduplication,
we assign each item a unique cluster id offline. Given the scale of over ten million items, as is shown in Figure \ref{fig:FBC}, we employ FAISS~\cite{douze2024faiss} for efficient offline similarity search and clustering. 

\subsubsection{Adapting to FAISS data processing}
We apply dimension-specific weights $ w^k $ to the corresponding embeddings $ \mathbf{e}^k $, producing weighted embeddings for the $ D-1 $ dimensions (excluding quantity).


The item quantity $ q $ is treated separately and mapped nonlinearly to a point $ (\cos\theta, \sin\theta) $ on the unit circle's first quadrant:

\begin{equation}
    \theta=\frac{\pi}{2} \times\left(1-e^{{-\alpha(q-1)}/{Q}}\right)
\end{equation}

where $ \alpha $ controls the nonlinearity of the mapping and $ Q $ is the maximum possible value of $ q $. This ensures that angular distance between points reflects semantic similarity in quantity.

Finally, we construct the unified item representation $ \mathbf{x}$:

\begin{equation}
    \mathbf{x} = \left[ \sqrt{w^1} \mathbf{e}^1, \dots, \sqrt{w^{D-1}} \mathbf{e}^{D-1} ,\, \sqrt{w^D} \cos\theta,\, \sqrt{w^D} \sin\theta \right]^\top.
\end{equation}

where $w^D$ is the weight of item quantity $q$.

\subsubsection{FAISS-Based Clustering}
Based on the processed item embeddings, we perform FAISS-based retrieval for each item $I_i$, returning all items within an $L_2$ distance threshold $\tau$:

\begin{equation}
\begin{split}
    \mathcal{C}_i &= \{ I_j \mid d_{i,j} < \tau, \ j \in \{1, \dots, N\} \} \\
    &= \mathrm{FAISS\_search}(I_i, \mathcal{I}, \tau), \quad \forall i \in \{1, \dots, N\}
\end{split}
\end{equation}


where $d_{i,j} = 2(1 - \mathbf{x}_i \cdot \mathbf{x}_j)$ denotes the $L_2$ distance between embeddings, and $N$ is the total number of items. Due to potential cluster overlaps, we apply a deduplication step to assign each item a unique cluster ID, resulting in $n$ final non-overlapping clusters.

\section{EXPERIMENTS}

To validate the effectiveness of \textbf{FITRep}, we compare it against a black-box method (\textbf{BBM}), which generate multimodal summaries through simple textual prompts, encode them using BERT, and apply dimensionality reduction. We exclude NoteLLM-2~\cite{zhang2025notellm} from our main comparison, as it is designed to incorporate cooperative signals through fine-tuning, whereas our goal is to learn intrinsic semantic representations independent of cooperative signals. We treat \textit{item deduplication} (including \textit{duplicate item identification} and \textit{duplicate item removal}) as the primary task, with CTR prediction serving as an indirect validation of embedding quality, assuming that semantically meaningful representations improve item-user matching. 

\subsection{Duplicate Item Identification}
For \textit{duplicate item identification}, we evaluate on a manually annotated dataset, defining positive pairs as duplicates and negative pairs as non-duplicates.

\begin{table}[h]
\centering
\caption{Offline Item deduplication performance}
\label{tab:item_deduplicatio}
\begin{tabular}{lccc}
\toprule
\textbf{Methods} & \textbf{Precise} & \textbf{Recall}  & \textbf{F1}\\
\midrule
\textbf{BBM} & 56.9\% & \textbf{95.0\%} & 71.2\% \\
\textbf{FITRep}  & \textbf{88.1\%} & 87.5\% & \textbf{87.8\%} \\
\bottomrule
\end{tabular}
\end{table}

As shown in Table~\ref{tab:item_deduplicatio}, \textbf{FITRep} significantly improves precision (88.1\% vs. 56.9\%) over \textbf{BBM}, with only a small drop in recall (87.5\% vs. 95.0\%), resulting in a much higher F1-score (87.8\% vs. 71.2\%). This demonstrates that attention-guided fine-grained embeddings enable more accurate duplicate detection.

\subsection{Impact of Dimensionality Reduction Methods}
On manually annotated dataset, we evaluate the impact of various dimensionality reduction methods. As shown in Table~\ref{tab:performance-dr}, \textbf{FITRep}, leveraging parameterized UMAP, surpasses both PCA-based and VAE-based methods, achieving the highest precision (88.1\%), recall (87.5\%), and F1-score (87.8\%), demonstrating its superior ability to preserve structure during dimensionality reduction.

\begin{table}[ht]
    \centering
    \caption{Comparison of Dimensionality Reduction Methods.}
    \label{tab:performance-dr}
    \begin{tabular}{lccc}
    \toprule
    \textbf{Methods} & \textbf{Precise} & \textbf{Recall}  & \textbf{F1}\\
    \midrule
    PCA-based & 75.3\% & 81.4\% & 78.2\% \\
    VAE-based & 81.2\% & 85.7\% & 83.4\% \\
    \textbf{FITRep}  & \textbf{88.1\%} & \textbf{87.5\%} & \textbf{87.8\%} \\
    \bottomrule
    \end{tabular}
\end{table}

\subsection{CTR Prediction}

We evaluate our framework on CTR prediction using a large-scale industrial dataset from Meituan's advertising system, which captures real user interactions in a massive local-services ecosystem. Key statistics are summarized in Table~\ref{tab:dataset_statistics}.

\begin{table}[h]
\centering
\caption{Dataset statistics.}
\label{tab:dataset_statistics}
\begin{tabular}{lccc}
\toprule
\textbf{Dataset} & \textbf{\#Requests} & \textbf{\#Users} & \textbf{\#Items} \\
\midrule
Meituan & 98,362,548 & 24,215,750 & 5,365,286 \\
\bottomrule
\end{tabular}
\end{table}

Our proposed method, \textbf{CTR\_FITRep}, integrates item embeddings for CTR prediction, and is compared against \textbf{CTR\_BBM}, which uses item representations generated by \textbf{BBM} method.

\begin{table}[ht]
    \centering
    \caption{Offline performance on the Meituan dataset.}
    \label{tab:performance-comparison}
    \begin{tabular}{lcc}
        \toprule
        \textbf{Model} & \textbf{AUC} $\uparrow$ & \textbf{LogLoss} $\downarrow$ \\
        \midrule
        CTR\_BBM   & 0.6580 & 0.0234 \\
        \textbf{CTR\_FITRep} & \textbf{0.6640} & \textbf{0.0215} \\
        \bottomrule
    \end{tabular}
\end{table}

Following standard practice, we report \textbf{AUC}  and \textbf{LogLoss}. 
As shown in Table~\ref{tab:performance-comparison}, 
the gain over \textbf{CTR\_BBM} (+0.6pp in AUC) demonstrates that replacing black-box prompt summaries with white-box, attention-guided representations better captures item semantics, leading to more accurate user interest modeling. 

\subsection{Online Deployment and A/B Test Results}
We deploy \textbf{FITRep} on Meituan’s advertising system to support two key tasks: \textit{item deduplication} (primary) and \textit{CTR prediction} (auxiliary).

For efficient online \textit{item deduplication}, we leverage \textbf{FITRep} embeddings to cluster items during offline processing and store their cluster IDs in Redis. At serving time, for each request, we retrieve the cluster IDs of all candidate items, group those sharing the same cluster ID as duplicates, and retain only the highest-ranked item per cluster, yielding the \textbf{FITRep\_ID} strategy.

To further enhance CTR prediction, we integrate \textbf{FITRep}-based semantic representations into the ranking model (\textbf{CTR\_FITRep\_ID}). For comparison, we also evaluate a variant that uses FITRep embeddings without \textit{item deduplication} (\textbf{CTR\_FITRep}).

We conducted an online A/B test from October 6–14, 2025. As shown in Table~\ref{tab:ab_test_results}, \textbf{FITRep\_ID} improves user experience by eliminating redundant ads, achieving +2.15\% CTR and +2.40\% CPM over the Baseline with negligible latency overhead (21.5 ms vs. 21.1 ms). When combined with the CTR model, \textbf{CTR\_FITRep\_ID} delivers substantial gains of +3.60\% in CTR and +4.25\% in CPM, significantly outperforming both the Baseline and the non-deduplicated variant (\textbf{CTR\_FITRep}). 

\begin{table}[ht]
    \centering
    \caption{Online A/B testing results on Meituan's advertising system.}
    \label{tab:ab_test_results}
    \begin{tabular}{lccc}
        \toprule
        \textbf{Method} & \textbf{CTR Gain} & \textbf{CPM Gain} & \textbf{Latency} \\
        \midrule
        Baseline               & —      & —      & 21.1 ms \\
        \textbf{FITRep\_ID}                & +2.15\% & 2.40\% & 21.5 ms \\
        \midrule
        CTR\_FITRep            & +1.50\% & +1.68\% & 21.3 ms \\
 \textbf{CTR\_FITRep\_ID} & \textbf{+3.60\%} & \textbf{+4.25\%} & 21.8 ms \\
        \bottomrule
    \end{tabular}
\end{table}

\section{Conclusion}
We identify the \textit{local structural collapse} problem in existing black-box multimodal representations, which leads to false-positive duplicates and degraded user experience. To address this, we propose \textbf{FITRep}, a white-box, attention-guided framework inspired by FIT that enables fine-grained, interpretable item representation through concept-hierarchical extraction, structure-preserving dimensionality reduction, and scalable indexing. Deployed on Meituan's advertising system, \textbf{FITRep}  significantly improves user engagement, yielding +3.60\% CTR and +4.25\% CPM in online A/B tests, demonstrating its practical value in large-scale recommendation systems.

\bibliographystyle{ACM-Reference-Format}
\balance
\bibliography{mllm.bib}
\end{document}